\documentclass{article}

\usepackage{PRIMEarxiv}

\usepackage[utf8]{inputenc} % allow utf-8 input
\usepackage[T1]{fontenc}    % use 8-bit T1 fonts
\usepackage[hidelinks,linkcolor=black]{hyperref}
\usepackage{url}            % simple URL typesetting
\usepackage{booktabs}       % professional-quality tables
\usepackage{amsfonts,amsmath,amssymb}       % blackboard math symbols
\usepackage{nicefrac}       % compact symbols for 1/2, etc.
\usepackage{microtype}      % microtypography
\usepackage{lipsum}
\usepackage{fancyhdr}       % header
\usepackage{graphicx}       % graphics
\graphicspath{{media/}}     % organize your images and other figures under media/ folder

\usepackage{subfigure}
\title{Dynamic Pricing for Electric Vehicle Charging
}

\author{
  Arun Kumar Kalakanti\\
  Toshiba Software (India) Pvt. Ltd. R\&D \\
  Bengaluru, India\\
  \texttt{arun.kalakanti@iiitb.ac.in} \\
   \And
  Shrisha Rao \\
  International Institute of Information Technology Bangalore\\
  Bengaluru, India\\
  \texttt{shrao@ieee.org} \\
}

\begin{document}
\maketitle

\begin{abstract}
Dynamic pricing is a promising strategy to address the challenges of smart charging, as traditional time-of-use (ToU) rates and stationary pricing (SP) do not dynamically react to changes in operating conditions, reducing revenue for charging station (CS) vendors and affecting grid stability.  Previous studies evaluated single objectives or linear combinations of objectives for EV CS pricing solutions, simplifying trade-offs and preferences among objectives. We develop a novel formulation for the dynamic pricing problem by addressing multiple conflicting objectives efficiently instead of solely focusing on one objective or metric, as in earlier works. We find optimal trade-offs or Pareto solutions efficiently using Non-dominated Sorting Genetic Algorithms (NSGA) II and NSGA III. A dynamic pricing model quantifies the relationship between demand and price while simultaneously solving multiple conflicting objectives, such as revenue, quality of service  (QoS), and peak-to-average ratios (PAR).  A single method can only address some of the above aspects of dynamic pricing comprehensively.  We present a three-part dynamic pricing approach using a Bayesian model, multi-objective optimization, and multi-criteria decision-making (MCDM) using pseudo-weight vectors. To address the research gap in CS pricing, our method selects solutions using revenue, QoS, and PAR metrics simultaneously. Two California charging sites’ real-world data validates our approach.
\end{abstract}

% keywords can be removed

\section{INTRODUCTION}
Electric vehicles (EVs) have gained popularity due to their cost- and eco-effectiveness~\cite{Moghaddam2018}. According to a projection from the International Energy Agency~\cite{ie2019}, the worldwide EV fleet is expected to reach 250 million by 2030. A widespread public EV charging infrastructure is necessary for the electrification of transportation, in addition to the proper integration of EVs into the power grid. Smart charging~\cite{wang2016_price}, which enables the coordinated regulation of charging operations, is considered an important step towards the proper grid integration of EVs and the efficient operation of public EV charging stations (CSs). Dynamic pricing is a promising strategy to deal with the smart charging problems brought on by the growing use of EVs~\cite{en12183574,Limmer2019}. 

Dynamic pricing is a real-time pricing strategy based on factors like demand, supply, time, location, and customer behavior. Static or Stationary pricing (SP) is fixed over time and may not be suitable for contexts like EV charging with variable demand. Dynamic pricing in EV charging manages demand, avoids congestion, incentivizes off-peak charging, and promotes efficient infrastructure use. It has potential benefits such as lower energy production costs, increased grid stability, higher user satisfaction, and lower operating costs for charging stations. This pricing approach has gained significant research attention in the EV charging industry~\cite{Limmer2019,Moghaddam2019,Xiong2016,Wolbertus2018}. However, past research has not considered multiple metrics and planning areas for the investigation, which are important in practical settings.

Dynamic pricing enables businesses to adjust prices in response to changing market conditions, consumer demand patterns, and price competition in industries like transportation (e.g., ride-sharing apps), hospitality (e.g., hotel bookings), e-commerce (e.g., online marketplaces), and entertainment (e.g., ticketing). This helps businesses increase revenue and profit. However, there are unique challenges in EV charging, such as managing demand, preventing congestion, incentivizing off-peak charging, addressing range anxiety, and ensuring transparency in pricing. Effective implementation of dynamic pricing in EV charging requires careful consideration of these specific issues to ensure efficient utilization of charging infrastructure and a positive charging experience for EV users.

Time-of-use (ToU) rates are commonly utilised for pricing. However, they primarily take into account the duration of use and do not dynamically react to changes in operating conditions~\cite{Zeng2008,Nicolson2018}. The inefficiency of ToU can be attributed to several factors: (a) ToU pricing does not always take into account the quality of service provided at different times, (b) many industries have demand patterns that cannot be easily aligned with predefined ToU periods, and (c) public transit systems and operating equipment require continuous energy use, so they may not be suitable for ToU pricing because operations cannot be easily shifted to off-peak hours.

\begin{table}
\centering
\caption{Dynamic Pricing Solutions for EV Charging} 
\label{dpsol} 
\setlength\tabcolsep{0.3pt}
\

\begin{tabular*}{\textwidth}{@{\extracolsep{\fill}}l*{11}{l}}
\toprule
\hline
Reference (Year) & Objectives & Approach & Metric & Baseline\\
\hline
\addlinespace

Xu~\textit{et al.}~\cite{Xu2017}(2017) & Balance load & Heuristics & Revenue and  & Stationary Pricing\\
&&&Travel time to CS&& \\
Xu~\textit{et al.}~\cite{Xu2018}(2018) & Min Waiting & Heuristics & QoS and  & Stationary Pricing \\
& time &&Waiting time&& \\
Moghaddam~\textit{et al.}~\cite{Moghaddam2020} & Balance load & RL & Revenue and & Coordinated Dynamic Pricing\\
(2020) & & &Energy saving &  \\
% Xu~\textit{et al.}~\cite{Xu2020}(2020) & Max Revenue and Min load & NSGA II & Load value and Charging power & Uncoordinated Charging price\\
Xu~\textit{et al.}~\cite{Xu2020}(2020) & Max Revenue & NSGA II & Load value and & Uncoordinated Charging price\\
& and Min load & & Charging power \\
Fang~\textit{et al.}~\cite{Fang2021}(2021) & Max Revenue & Reinforcement & Revenue & Stationary Pricing \\
&&Learning (RL)&&& \\
Zhang~\textit{et al.}~\cite{Zhang2022}(2022) & Max Revenue & RL & Revenue & ToU\\
Liu~\textit{et al.}~\cite{Liu2023}(2023) & Balance load & Genetic Algorithm & Load value & Base grid load\\ 
\hline
\bottomrule
\hline
\end{tabular*}
\end{table}

In the literature, many strategies for maximizing EV charging station operators' profits~\cite{Mehta2016,Goebel2016,Naharudinsyah2018} and boosting grid stability~\cite{Lopes2009,Waraich2013} are suggested. Individual users may also use smart charging to lower their charging expenses~\cite{Rotering2011,Iversen2014,Kontou2017}. In recent years, a variety of innovative approaches to dynamic pricing in the context of EV charging have been developed and published~\cite{Limmer2019,Moghaddam2019,Xiong2016,Wolbertus2018}. 
However, the majority of current dynamic pricing research assumes only one objective~\cite{Xu2017,Xu2018,Fang2021,Aung2021,Liu2023} or that multiple objectives can be adequately handled through a simple linear combination of objectives~\cite{Xu2020, Moghaddam2020,Zhang2022} as shown in Table~\ref{dpsol}. It is inefficient to solve multi-objective problems linearly because it oversimplifies trade-offs and preferences among objectives, resulting in sub-optimal solutions that fail to reflect the complexity of real-world decision-making processes~\cite{Siirola2003, coello2006}. Furthermore, it assumes fixed weightings for objectives, which do not accurately reflect changing preferences or priorities. This work solves Multi-objective optimization efficiently by finding a set of trade-off optimal solutions, known as Pareto-optimal solutions, using Pareto-based optimization methods Non-dominated Sorting Genetic Algorithm (NSGA) II and NSGA III. These Pareto solutions are shown to solve complex problems by finding optimal solutions for conflicting objectives~\cite{Zitzler1998,Deb2011}.

Dynamic pricing involves quantifying the relationship between price and demand as well as solving multiple-objective problems involving revenue, quality of service (QoS), and peak-to-average ratios (PAR). In order to solve dynamic pricing comprehensively, a single method is unlikely to address all aspects of the problem~\cite{Deb2011}. With dynamic pricing, multiple sub-problems must be solved efficiently using a single method, including prime demand quantification and multi-objective revenue, quality of service, and PAR optimization. Multifaceted challenges often require a combination of strategies and algorithms tailored to each sub-problem~\cite{Zitzler1998}. Additionally, certain methods are more appropriate for addressing specific sub-problems. For example, Bayesian models are better suited to quantifying demand prices, while NSGA variants are better suited to solving multi-objective problems. Also, using a single method like ToU to solve all aspects of dynamic pricing could result in sub-optimal results. In this work, we have broken down the dynamic pricing problem into three separate problems: (1) The automatic quantification of demand using a Bayesian model; (2) the Multi-objective optimization of revenue, QoS, and PAR for a given price and demand vectors; and (3) Automated selection of balanced solutions using multi-criteria decision-making (MCDM).

In previous studies, evaluation of EV CS pricing solutions has been done predominantly using a single metric like revenue~\cite{Limmer2019}, QoS~\cite{Wolbertus2018}, and PAR~\cite{Moghaddam2019}. Revenue, QoS, and PAR metrics are important considerations, but they are not taken together for several reasons: (a) optimizing multiple metrics simultaneously can result in significant optimization problem complexity, which may make finding efficient and practical solutions more challenging. (b) managing peak loads or maintaining high-quality service can conflict with revenue maximization, which may require difficult trade-offs, and (3) algorithms and systems designed to maximize multiple metrics simultaneously while capturing real-world complexities can be hard to create and require advanced optimization techniques. Moreover, limited investigations have been done which consider the real charging data in multiple planning areas. Besides, previous studies have not dealt with the MCDM problem to select solutions automatically, given the importance of metrics. Also, an analysis of the automation of the pricing problem and the solution-finding capabilities considering these important parameters has not been undertaken. In this work, we created a dynamic pricing system to ensure a well-balanced optimal solution when key metrics like revenue, QoS, and PAR are considered. Real productivity benefits may result from focusing on such a pricing strategy.

In this work, an investigation has been done by considering the real charging session data in multiple planning areas in California using ACN Open EV Charging Dataset~\cite{ACNdata}. Caltech site contains 54 EV chargers and a 50 kW DC fast charger on the campus. In a parking garage, 52 EV chargers are part of JPL's site. Unlike Caltech, the JPL campus has gated access, and only workers are permitted to use the charging station. Caltech is a cross between workplace and public use charging, whereas the JPL site is an example of workplace charging. EV penetration is also relatively high on the JPL site.

Overall, the key contributions of this paper are as follows:
\begin{enumerate}
\item A novel problem formulation containing multiple optimization objectives of revenue, QoS, and PAR has been developed to solve the CS pricing problem.
\item An investigation of a three-part dynamic pricing method using a Bayesian model for demand-price quantification, multi-objective optimization using fast approaches~\cite{nsga2Deb,nsga3Deb} like Non-dominated Sorting Genetic Algorithm (NSGA) II and NSGA III, and MCDM with pseudo-weight vector methodology.
    \item A research gap in CS pricing has been addressed, viz., the exclusion of multiple objectives like revenue, QoS, and PAR while finding a pricing solution.
    \item The EV CS pricing solutions have been evaluated using real EV charging datasets in two planning areas: the JPL area, an example of workplace charging, and the Caltech site, a hybrid of workplace and public use charging.
    \item Another major contribution of this work is in identifying a pricing solution automatically using the pseudo-weight vector methodology of MCDM or if the importance of metrics is given. Additionally, there hasn't previously been a thorough investigation of the price problem's automation and the capacity for developing solutions while taking these metrics importance parameters into account.
\end{enumerate}

\section{PROBLEM DESCRIPTION AND FORMULATION}\label{problem}
Electric utility companies can influence the charging behavior of EV owners by manipulating the pricing of charging services at different CSs~\cite{wang2016_price}. The ``demand curve", a relationship between price and units required by consumers, is generally a factor that affects the cost of charging and revenue. Dynamic pricing optimization involves considering the demand curve, combining revenue generation with cost coverage, and adjusting prices to maximum profit. Along with revenue maximisation, challenges in pricing optimization for EV charging include decreasing the PAR load and improving the quality of service.

Three related problems are used to formulate the problem of dynamic price optimization: (1) Calculation or quantification of the demand-price relationship between EV and CS; (2) multi-objective optimization of three metrics (maximize revenue, minimize the difference between peak and average loads, and maximize the quality of service) using the obtained demand-price relationship to determine the price vector for various time slots, and (3) multi-criteria decision making to produce the best option by taking into account multiple criteria in the selection process.

\subsection{Demand-Price Model}
If $P_{\mathrm{CS}}$ and $D_{\mathrm{CS}}$ denote the CS pricing and the quantity of power demanded, the relationship between them is in terms of a conditional expectation of the form
\begin{equation}\label{eq1}
    \mathbf{E}[D_{CS}|P_{CS}] = a \times P_{CS}^{c}
\end{equation}
\begin{equation}\label{eq2}
    \log \mathbf{E}[D_{CS}|P_{CS}] = \log a + c \times \log P_{CS}
\end{equation}
$\mathbf{E}[D_{CS}|P_{CS}]$ represents the expected or average charging demand $D_{CS}$ at a specific price $P_{CS}$. The constant term $a$ represents the level of charging demand at the lowest price. As a measure of the sensitivity of demand to price variations, the elasticity coefficient $c$ is commonly referred to as the price elasticity of demand. Depending on the nature of the relationship, $c$ can have a positive, negative, or zero value. The values of $a$ and $c$ can be determined using empirical data. The demand at a CS for a given price $P_{CS}$, i.e., $D_{\mathrm{CS}}$ is obtained from demand-price quantification using a Bayesian model. This considers the fact that $D_{\mathrm{CS}}$ follows a distribution established by $P_{\mathrm{CS}}$ rather than being deterministic~\cite{Liu2021, Xu2020}. In log-log space, this linear relationship translates to constant price elasticity, as shown in~\eqref{eq2}. Since the elasticity is a constant and only depends on the parameter $c$, the elasticity can be estimated by fitting the model. 

\subsection{Multi-Objective Formulation}\label{mop}
The problem is modeled with multiple global optimization objectives. The fitness function $F$ comprises three objectives---the revenue (\(f_{\mathrm{revenue}}\)), the QoS  (\(f_{\mathrm{QoS}}\)), and the PAR (\(f_{\mathrm{PAR}}\)). Since pricing solutions for all CSs are obtained together by sharing their price and demand information, cooperation is present among CSs, and thus, they are complementary. 

The demand at a CS at time $t$ for a given price $P_{\mathrm{CS, t}}$, i.e., $D_{mean}({CS,t})$ is obtained from demand-price quantification using a Bayesian model. The sum of the revenue from a charging activity at a CS at time $t$ when the demand at the CS is $D_{mean}({CS,t})$ is referred to as \emph{revenue}. We calculate the total revenue of all CSs given a price vector $\mathbf{P}$ containing prices for all CSs and at all slots as below\\
\begin{equation}\label{eq3}
f_{\mathrm{revenue}}(\mathbf{P}) = \sum_{CS=1}^{N_{CS}}\sum_{t=1}^{T} P_{\mathrm{CS, t}} \times D_{mean}({CS,t})  
\end{equation}
where \(f_{\mathrm{revenue}}(.)\) calculates the total revenue for charging the price $P_{\mathrm{CS, t}}$ at a CS for all CSs and for all time slots. The pricing vector $\mathbf{P}$ assumes that different prices apply to each slot and that the users know the prices at each slot.

The highest power demand seen during a given time is known as the peak load of a system. The portion between the average load, denoted by $D_{mean}(CS,t)$, and the peak load, denoted by $D_{max}(CS,t)$, is defined as a system load factor. The load factor reveals if a system experiences significant demand changes. Large variations are a result of a low load factor. We compute the load factor referred to PAR as \\
\begin{equation}\label{eq4}
f_{\mathrm{PAR}}(\mathbf{P}) = \frac{1}{T \times N_{CS}} \sum_{CS=1}^{N_{CS}} \sum_{t=1}^{T} \frac{D_{max}(CS,t)}{D_{mean}(CS,t)} 
\end{equation}
We calculate the QoS of a CS at time $t$ as the ratio of the demand served at time $t$, $D_{delv}(CS,t)$, and the requested demand, $D_{req}(CS,t)$. We compute the average QoS of all CSs for a price $\mathbf{P}$ as\\
\begin{equation}\label{eq5}
f_{\mathrm{QoS}}(\mathbf{P}) = \frac{1}{T \times N_{CS}} \sum_{CS=1}^{N_{CS}} \sum_{t=1}^{T}
\dfrac{D_{delv}(CS,t)}{D_{req}(CS,t)} \\
\end{equation}

The mathematical formulation of the multi-objective problem is defined below:
\begin{eqnarray}
\textrm{max} \: f_{revenue}(\mathbf{P}) \label{eq6}\\
\textrm{min} \: f_{PAR}(\mathbf{P}) \label{eq7}\\
\textrm{max} \: f_{QoS}(\mathbf{P}) \label{eq8}\\
\textit{such that} \: 0 \leq P_{CS,t} \leq P_{max}  \label{eq9}\\
D_{delv}(CS,t) \leq CS_{capacity, t} \label{eq10}
\end{eqnarray}
where \(f_{\mathrm{revenue}}(.)\) calculates the total revenue at all CSs as shown in~\eqref{eq3}, $f_{PAR}(.)$ calculated the average PAR or load factor at all CSs as shown in~\eqref{eq4}, and $f_{QoS}(.)$ denotes the QoS of all CSs as shown in~\eqref{eq5}. The $P_{max}$ denotes the upper limit of a CS price, and the demand delivered $D_{delv, P_{CS,t}}$ is bounded by $CS_{capacity, t}$ which is the CS capacity at time $t$.

\subsection{Multi-Criteria Decision Making}
The set of non-dominated solutions where each objective is considered equally good is termed the Pareto-front. A Pareto-front with numerous alternative solutions is the output of a multi-objective algorithm. The problem is to choose a solution out of the solution set in the context of multi-objective optimization. MCDM problem refers to the problem of making decisions or selecting a solution that involves numerous criteria or objectives.  

\section{Dynamic Pricing Approach}
The dynamic pricing approach solves multiple sub-problems in three parts.
\subsection{Part I - Demand-Price Quantification using Bayesian Optimal Pricing}
Demand-price quantification refers to the process of estimating the relation between the power demand and its price to find the ideal price that will maximize revenue. 
We used Bayesian modeling for demand-price quantification to estimate the relationship between power demand and pricing based on historical charging data from the ACN Open EV Charging Dataset~\cite{ACNdata}. This statistical technique involves identifying a prior distribution based on existing knowledge and assumptions and then updating it with observed data using Bayes' theorem to determine the posterior distribution. The historical demand and price data from charging stations and the relationship shown in~\eqref{eq1} and~\eqref{eq2} are used for this quantification process.

The posterior distribution is used to predict the best price that will maximize revenue based on the observed data and our updated assumptions about the relationship between price and quantity desired. Bayesian modeling is utilized to quantify demand-price relationships, make accurate pricing strategy decisions, and increase the capacity to estimate demand and optimize revenue.

\subsection{Part II - Multi-Objective optimization using NSGA II or NSGA III}

Multi-objective optimization refers to the process of optimizing multiple conflicting objectives, such as~\eqref{eq6},~\eqref{eq7}, and~\eqref{eq8}, simultaneously with constraints such as~\eqref{eq9} and~\eqref{eq10}. NSGA (Non-dominated Sorting Genetic Algorithm) is a well-known and fast approach for solving multi-objective optimization problems~\cite{nsga2Deb,nsga3Deb}. We used two variations of the NSGA algorithm, i.e., NSGA-II and NSGA-III, for solving the multi-objective optimization problem described in Section~\ref{mop}. The two algorithms use a genetic algorithm technique, which entails producing a set of candidate solutions and then evaluating and developing these solutions to increase their fitness with respect to the numerous objectives.

The NSGA-II algorithm operates by assigning each candidate solution to a group of non-dominated solutions, which are solutions that are not dominated by any other solution in terms of multiple objectives, i.e., (a) maximize total CS revenue, \(f_{\mathrm{revenue}}(.)\) (b) maximize QoS, $f_{\mathrm{QoS}}$ and (b) minimize PAR, $f_{PAR}(.)$. The algorithm then selects a set of the best non-dominated solutions known as Pareto-front and generates a new set of candidate solutions based on these solutions.

NSGA-III is an expansion of NSGA-II that tries to increase its performance by employing a reference point technique to guide the search towards specified portions of the objective space~\cite{nsga3Deb}. In NSGA-III, a set of reference points is defined, and the algorithm searches for non-dominated solutions that are closest to these reference points.

\subsection{Part III - MCDM using pseudo weights}
We adopt the pseudo-weight vector methodology presented in~\cite{Blank2020} to select a solution from a solution set or Pareto-front in the context of multi-objective optimization. The objective behind pseudo weights is to give weights to the criteria based on their relative importance without forcing the decision maker to provide explicit numerical numbers for each criterion. This is done by performing pairwise comparisons between the criteria to assess their relative relevance. The pseudo weights $w_i$ for the objective $i$ is calculated as follows~\cite{Blank2020}:
\begin{equation}
    w_i = \frac{(f_i^{max} - f(x))/(f_i^{max} - f_i^{min})} {\sum_{m=1}^{M} (f_m^{max} - f_m(x))/(f_m^{max} - f_m^{min})}
\end{equation}

The objectives for different $i$ refer to \(f_{\mathrm{revenue}}(.)\), $f_{\mathrm{QoS}}$ and $f_{PAR}(.)$. The advantage of employing the technique of the pseudo weight is that it provides a mechanism to assess the relative relevance of each criterion without having explicit numerical values, which might be difficult or time-consuming to obtain. 

\section{RESULTS}
This section explains the evaluation setting and compares the performances of three solutions for dynamic pricing. 
\begin{table*}
\caption{Average Results of JPL and Caltech sites} 
\label{table1} 
\setlength\tabcolsep{0pt}
\
\begin{tabular*}{\textwidth}{@{\extracolsep{\fill}}l*{11}{r}}
\toprule
\hline
{} & {} & \bf JPL & {} & {} & \bf Caltech & {} \\
\addlinespace
\bf Pricing Approach/Metric & \bf Revenue(\$) & \bf QoS & \bf PAR & \bf Revenue(\$) & \bf QoS & \bf PAR \\
\hline
Stationary Price & 1613.64  & 0.34  & 16.77  & 5501.21  & 0.35  & 6.80  \\ 
ToU & 1691.87  & 0.42  & 16.08  &  6221.07  & 0.42  &  7.84 \\ 
%  & 11670.35 \pm 4711.79 & 0.40 \pm 0.07 & 15.97 \pm 8.22 & 101.68 \pm 57.17 & 0.55 \pm 0.11 & 9.48 \pm 6.53 \\
BM + NSGA II & 1880.46  & 0.41  & 4.89  & 6754.56  & 0.52  & 3.41  \\
BM + NSGA III & 1886.84  & 0.55  & 5.46  & 6584.22  & 0.60  & 5.40  \\
\bottomrule
\hline
\end{tabular*}
\label{AvgResult}
\end{table*}

\subsection{Evaluation Setting}
\begin{figure}[!htbp]
\centering
  \includegraphics[width=0.6\linewidth]{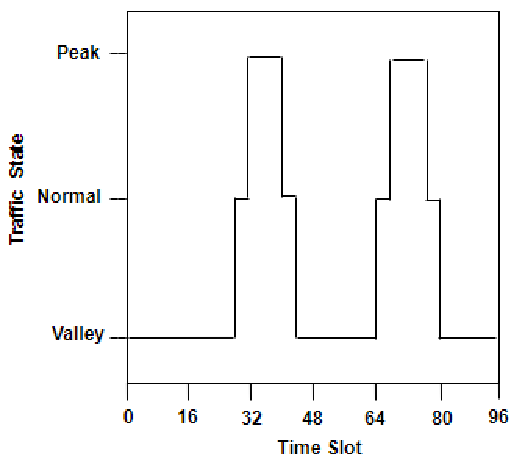}
  \caption{Time Slot}
  \label{ts}
\end{figure}

\begin{figure}[htbp]
\centering
  \includegraphics[width=0.8\linewidth]{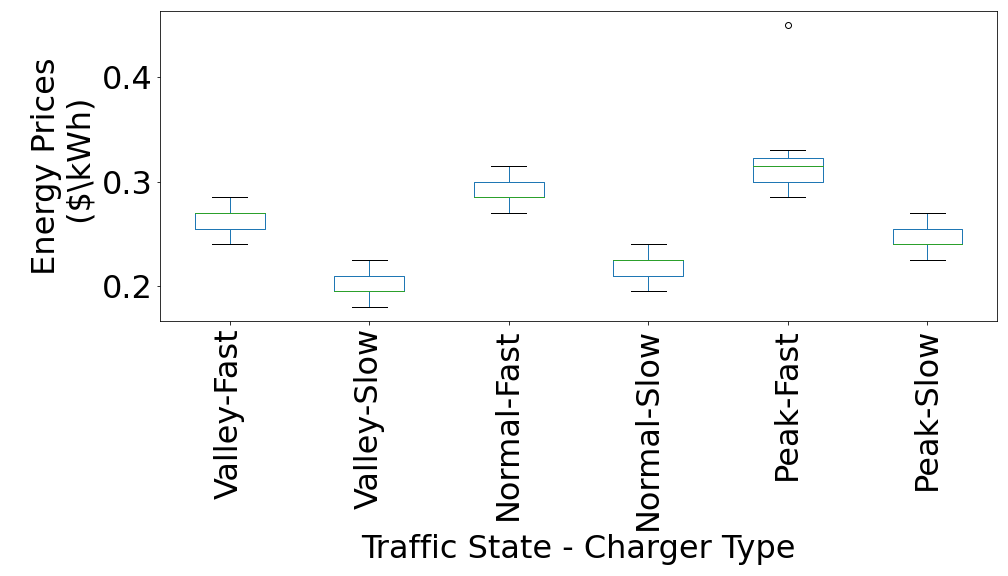}
  \caption{ToU Energy Prices}
  \label{gCS}
\end{figure}

Figure~\ref{ts} shows that one day has been divided into 96 time slots.  
There are two traffic peak slots around time slots 34 and 72 in a day. The box plot information containing the distribution of energy prices for the fast and slow charging options at different traffic states has been shown in Figure~\ref{gCS}.

Stationary pricing data is based on real-time site data. ToU prices are obtained using traffic state data from Figure~\ref{ts} and energy prices from Figure~\ref{gCS}. The quality of the developed solutions for all CSs is evaluated by three measures defined in~\eqref{eq2}: total revenue~\eqref{eq3}, QoS~\eqref{eq5}, and PAR~\eqref{eq4}. 

\subsection{Comparative Performance of All Approaches}
\begin{figure}[t]

\centering
\subfigure[Revenue]{
\includegraphics[width=.33\textwidth]{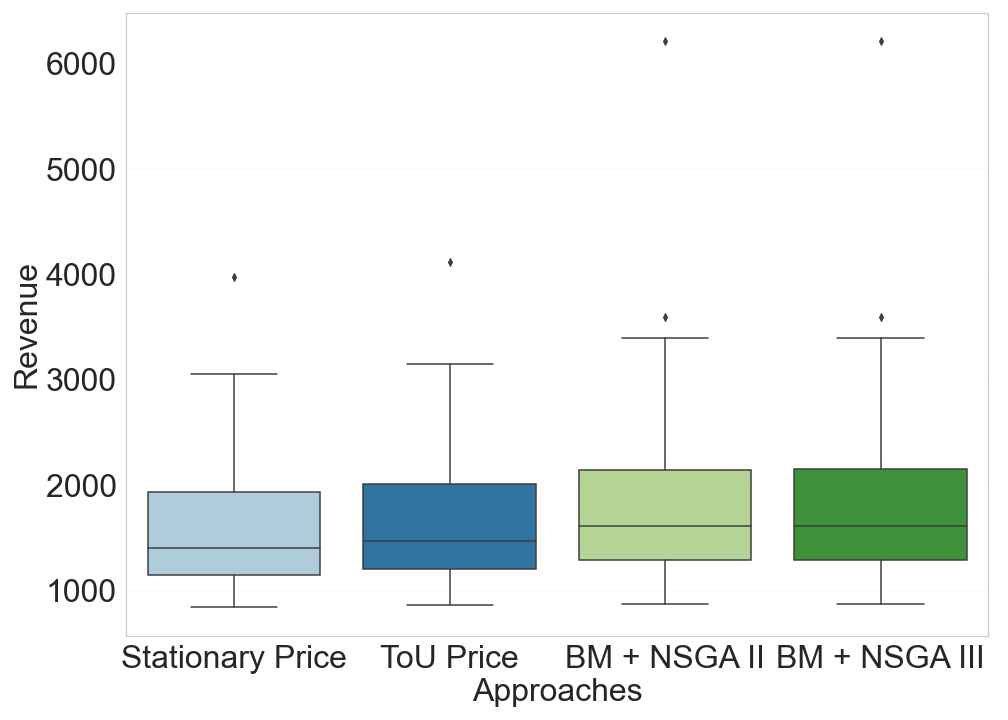}
}
\label{JPLRevenue}
\subfigure[QoS]{
\includegraphics[width=.33\textwidth]{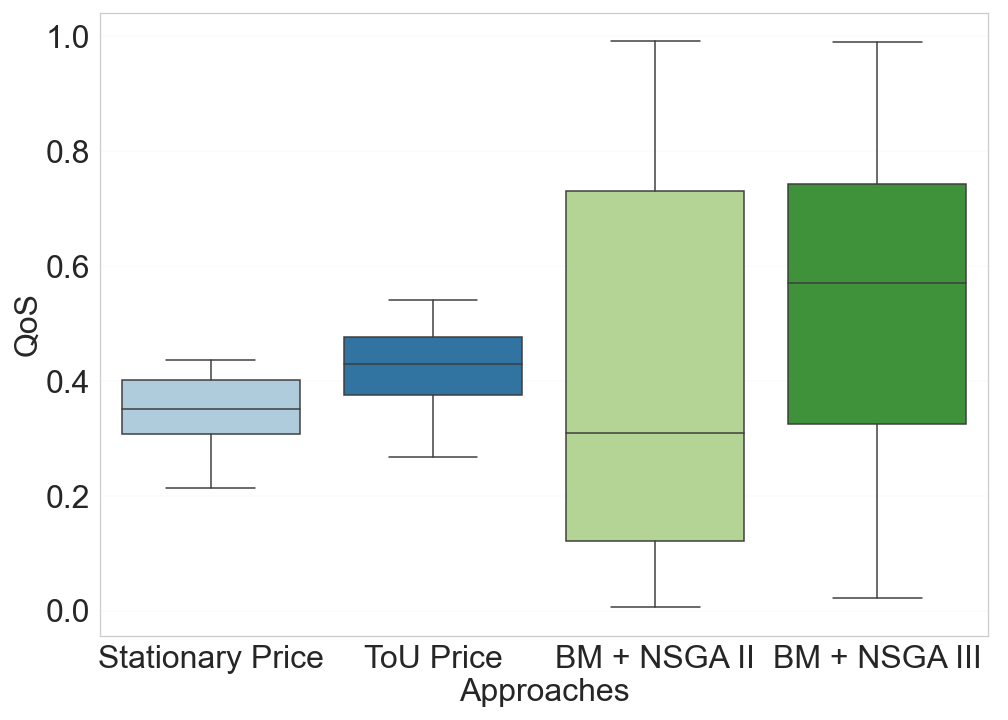}
}
\label{JPLQoS}
\subfigure[PAR]{
\includegraphics[width=.33\textwidth]{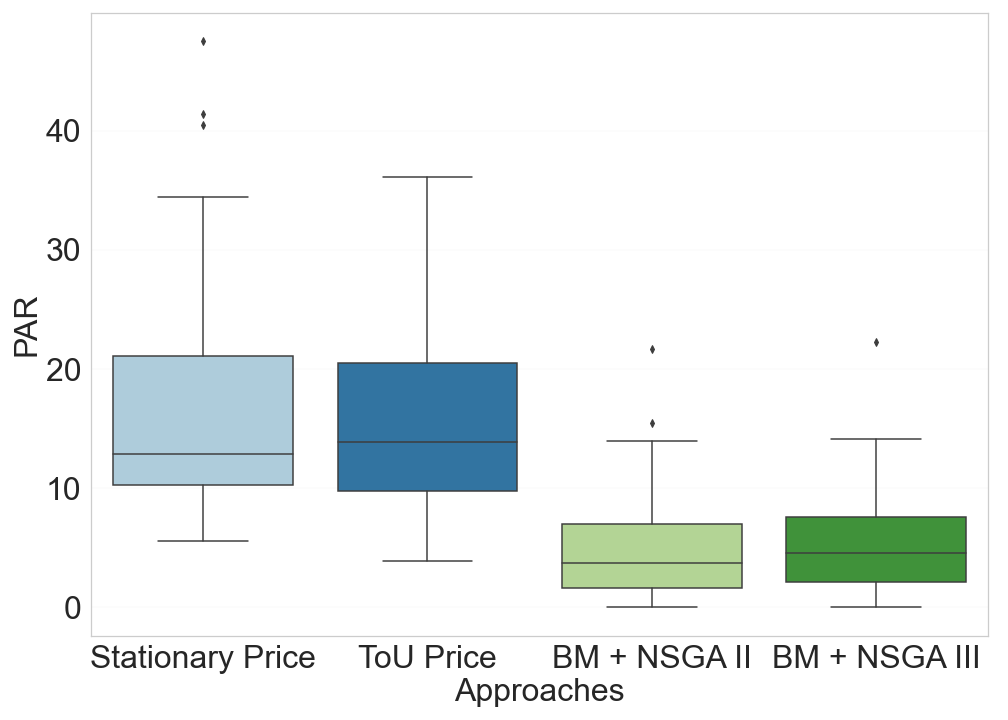}
}
\label{JPLPAR}
\caption{JPL Metrics}
\label{JPL}
\end{figure}

The average results of the three approaches are obtained and shown in Table~\ref{table1}. BM + NSGA II and BM + NSGA III refer to the solution approaches that use the Bayesian model in the first part of the solution, followed by NSGA II and NSGA III in the second part, respectively. MCDM is used in the third part of the solution for both approaches. BM + NSGA II and BM + NSGA III algorithms are compared with well-known ToU pricing and SP data for two sites, i.e., JPL and Caltech. 

\begin{table*}
\caption{Average Improvement Percentage at JPL site}  
\setlength\tabcolsep{0pt}
\
\begin{tabular*}{\textwidth}{@{\extracolsep{\fill}}l*{11}{c}}
\toprule
\hline
{} & {} & \bf Over SP & {} & {} & \bf Over ToU & {}\\
\addlinespace
\bf Pricing Approach/Metric & \bf Revenue(\$) & \bf QoS & \bf PAR & \bf Revenue(\$) & \bf QoS & \bf PAR \\
\hline

ToU & 4.85\% & 23.50\% & 4.11\% & - & - & -\\ 
BM + NSGA II & 16.54\% & 20.59\% & 70.84\% & 11.15\% & -2.38\% & 69.59\%\\
BM + NSGA III & 16.93\% & 61.76\% & 67.44\% & 11.52\% & 30.95\% & 66.04\%\\
\bottomrule
\hline
\end{tabular*}
\label{JPLPerc}
\end{table*}

\begin{table*}
\caption{Average Improvement Percentage at Caltech site} 
\setlength\tabcolsep{0pt}
\
\begin{tabular*}{\textwidth}{@{\extracolsep{\fill}}l*{11}{c}}
\toprule
\hline
{} & {} & \bf Over SP & {} & {} & \bf Over ToU & {}\\
\addlinespace
\bf Pricing Approach/Metric & \bf Revenue(\$) & \bf QoS & \bf PAR & \bf Revenue(\$) & \bf QoS & \bf PAR \\
\hline

ToU & 13.09\% & 20.00\% & 15.29\% & - & - & - \\ 
BM + NSGA II & 22.78\% & 48.57\% & 49.85\% & 8.58\% & 23.81\% & 56.51\% \\
BM + NSGA III & 19.69\% & 71.43\% & 20.59\% & 5.84\% & 42.86\% & 31.12\% \\
\bottomrule
\hline
\end{tabular*}
\label{CaltechPerc}
\end{table*}

\subsubsection{JPL site}

\paragraph{Revenue} Table~\ref{JPLPerc} shows the improvement of ToU over SP data, as well as the improvement of BM + NSGA II and BM + NSGA III variants over the base ToU and SP at JPL. BM + NSGA III algorithm improves the total revenue by 16.93\% compared to SP and 11.52\% compared to ToU. The revenue improvement of the BM + NSGA II algorithm is similar to that of the BM + NSGA III algorithm, and the differences between NSGA variants are insignificant. Overall, revenue is significantly improved with the three-part pricing algorithms compared to ToU and SP.

\paragraph{QoS} For the QoS at the JPL site, BM + NSGA III is the highest, i.e., the best performing, as shown in Table~\ref{JPLPerc} with an average improvement of 61.76\% over SP and 30.95\% over ToU. The average QoS using BM + NSGA II is slightly lower than ToU but better than SP. Figure~\ref{JPL}b shows that BM + NSGA III has a better QoS box plot than BM + NSGA II, and traditional methods like SP and ToU have fewer deviations.

\paragraph{PAR} BM + NSGA II and BM + NSGA III performances are similar for PAR as shown in Figure~\ref{JPL}c. However, the percentage improvement of BM + NSGA II is slightly better, as shown in Table~\ref{JPLPerc}. Also, both approaches reduced the PAR by 66-71\% compared to SP and ToU, as shown in Table~\ref{JPLPerc}. 

\begin{figure}[thbp!]
\centering
\subfigure[Revenue]{
\includegraphics[width=.33\textwidth]{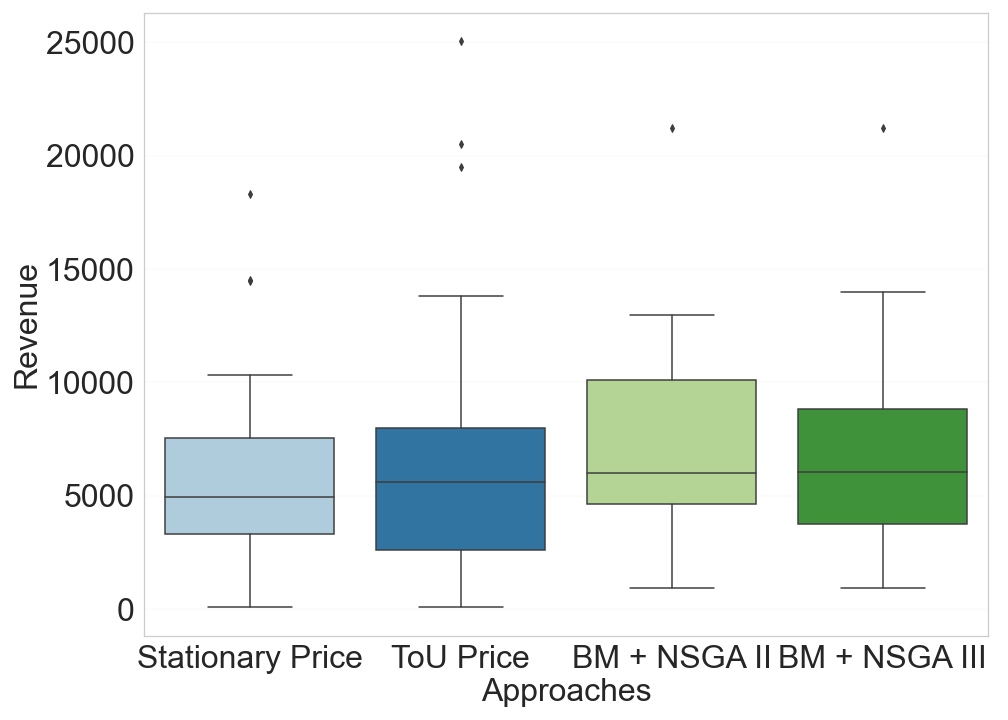}
}
\label{CRevenue}
\subfigure[QoS]{
\includegraphics[width=.33\textwidth]{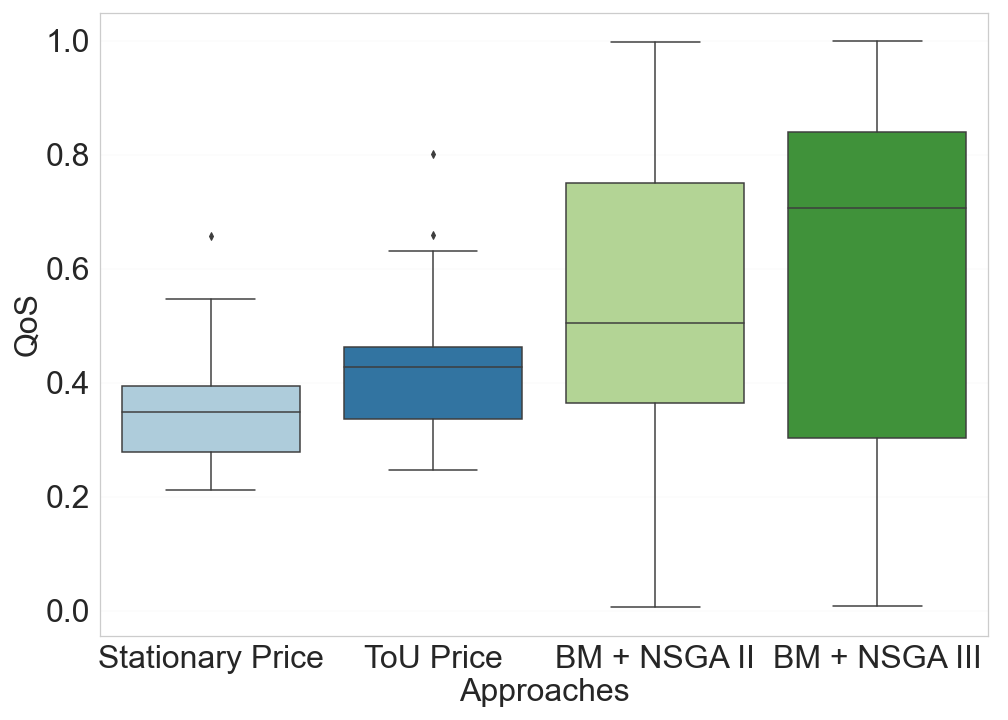}
}
\label{CQoS}
\subfigure[PAR]{
\includegraphics[width=.33\textwidth]{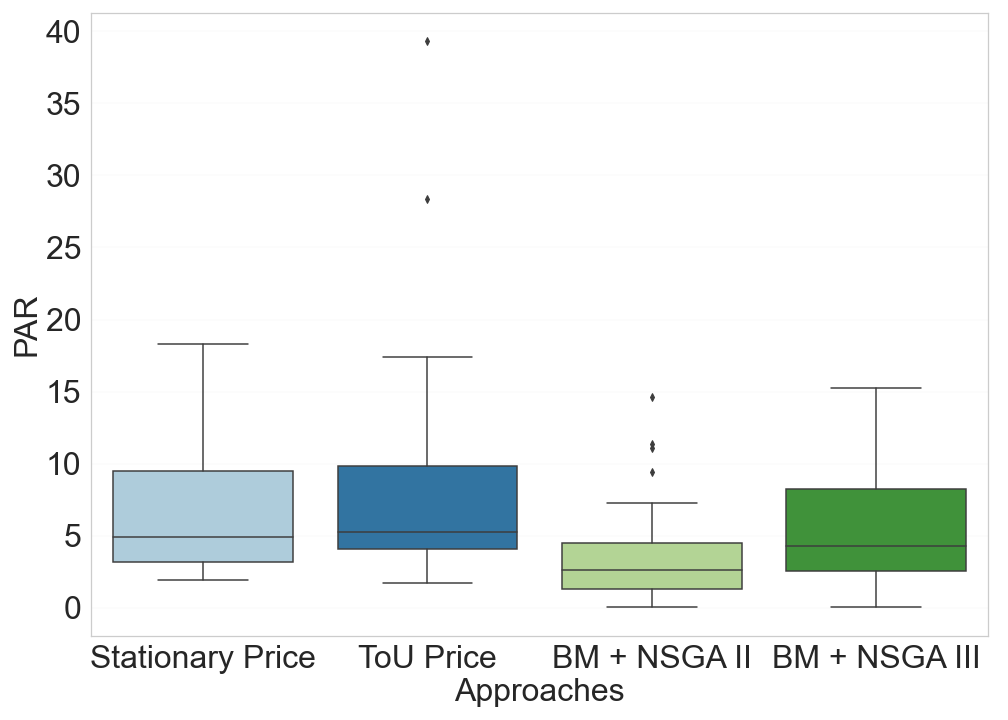}
}
\label{CPAR}
\caption{Caltech Metrics}
\label{Caltech}
\end{figure}

\subsubsection{Caltech site}

\paragraph{Revenue} Table~\ref{CaltechPerc} shows the improvement of ToU over SP data, as well as the improvement of BM + NSGA II and BM + NSGA III variants over the base ToU and SP at Caltech. BM + NSGA II algorithm improves the total revenue by 22.78\% compared to SP and 8.58\% compared to ToU. The revenue improvement of the BM + NSGA III algorithm is slightly lower than the BM + NSGA II algorithm. Overall, there is a significant improvement in revenue with the three-part pricing algorithms compared to ToU and SP.

\paragraph{QoS} For the QoS at the Caltech site, BM + NSGA III is the highest, i.e., the best performing, as shown in Table~\ref{CaltechPerc} with an average improvement of 71.43\% over SP and 42.86\% over ToU. The average QoS improvement using BM + NSGA II is lower than BM + NSGA III but better than ToU. Figure~\ref{Caltech}b shows that BM + NSGA III has a better QoS box plot than BM + NSGA II, and traditional methods like SP and ToU have fewer deviations.

\paragraph{PAR} The performance of BM + NSGA II is better for PAR as shown in Table~\ref{CaltechPerc} and Figure~\ref{CPAR}, and it improved the PAR by 48.85\% and 56.51\% compared to SP and ToU respectively.  

\subsubsection{Summary}
Table~\ref{table1} shows that BM + NSGA II and BM + NSGA III algorithms perform similarly in revenue for the JPL site. However, BM + NSGA II generates more revenue for Caltech. For QoS, BM + NSGA III performs best, i.e., highest at both sites. In contrast, BM + NSGA II has the lowest PAR, i.e., the best performing at both sites.  Overall, the BM + NSGA III algorithm is the best-performing algorithm for the JPL site due to improvement in most metrics compared to ToU and SP. Similarly, the BM + NSGA II algorithm is the best-performing algorithm at Caltech.

\subsection{Effect of selecting pricing solutions from Pareto-front using MCDM and importance factor}
The importance of metrics such as revenue or PAR reduction for large grids or microgrids will depend on the grid operators' and stakeholders' specific objectives and goals. If the primary objective is to maximize revenue, then revenue would be the most important metric. However, reducing PAR demand may be more important if the focus is on maintaining grid stability and reliability. This is because reducing peak demand can help prevent grid overloading, which can lead to power outages and other problems. In general, a balanced approach is recommended that considers both revenue and PAR reduction.

Two evaluation cases are further considered for the best-performing algorithms:(1) maximize revenue as an important factor for microgrids designed to operate as a commercial enterprise, and (2) PAR is an important metric for maintaining large grids stability and reliability. In the first case, the primary goal is to generate revenue by selling power to customers or participating in demand response programs. The second case deals with selecting solutions to reduce peak demand to prevent grid overloading, which can lead to power outages and other problems. JPL site is considered for this evaluation as the EV penetration is high in this case and the Pareto-front solutions obtained are high in number after Part II multi-objective optimization.

\begin{figure}[htpb]
  \centering
  \includegraphics[width=.48\textwidth]{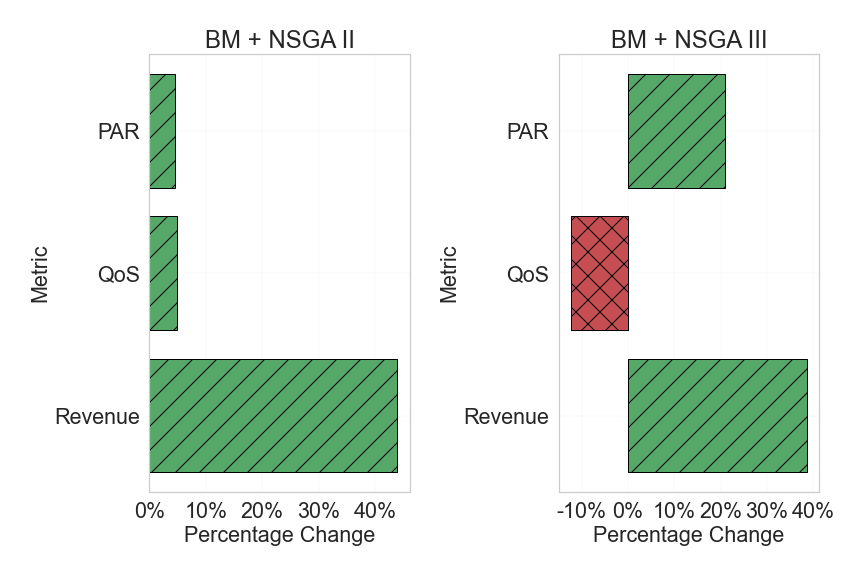}
  \caption{JPL with revenue metric as an importance factor}
  \label{MCDMJPLRev}
\end{figure}

\begin{figure}[htbp]
  \centering
  \includegraphics[width=.48\textwidth]{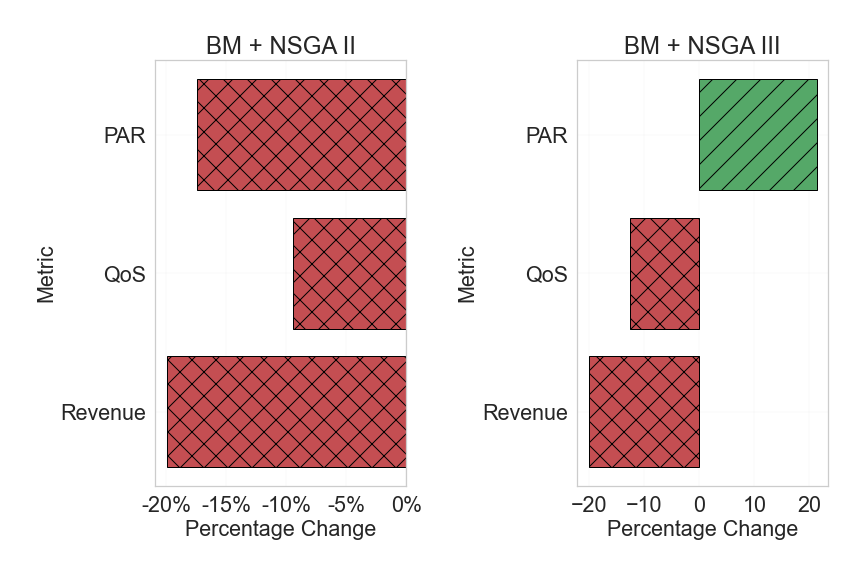}
  \caption{JPL with PAR metric as an importance factor}
  \label{MCDMJPLPAR}
\end{figure}

The BM + NSGA II and BM + NSGA III algorithms have numerous solutions, and the best solution is obtained by the MCDM method by computing pseudo weights. However, when some explicit weights are provided for the three metrics, a solution among the obtained solutions is selected. The resulting solutions of the MCDM, when high priority is given to revenue or the PAR metric and the subsequent improvement in the metrics, are presented in Figures~\ref{MCDMJPLRev} and~\ref{MCDMJPLPAR}.

Figures~\ref{MCDMJPLRev} show a steep increase, i.e., an improvement of 40\% in total revenue from earlier solutions obtained from BM + NSGA II and BM + NSGA III algorithms.  For the PAR, BM + NSGA III is the highest, i.e., the best performing compared to the improvement with BM + NSGA II solutions. However, BM + NSGA III algorithm compromised on QoS for about 10\% and BM + NSGA II slightly improved QoS.

Figures~\ref{MCDMJPLPAR} show that only BM + NSGA III could find the solution with PAR as an important metric with an improvement of 20\% in PAR. For the PAR, BM + NSGA II couldn't find a related solution, as can be seen by the PAR metric. However, total revenue and QoS from new solutions obtained from both BM + NSGA II and BM + NSGA III algorithms decreased significantly by 20\% and 10\% respectively.

Overall, the best-performing method that tried to provide a solution from the set of solutions with the given importance of the metric is BM + NSGA III compared to BM + NSGA II. This can be attributed to the fact that NSGA III improves the diversity of the obtained solutions by using reference points, which helps to guide the search towards a diverse set of non-dominated solutions. Also, NSGA III utilizes a new selection operator, known as the orthogonal selection operator, which enhances the diversity of the offspring population. The NSGA III algorithm used an enhanced diversity metric to measure the spread of solutions in the population. It was able to find a wide range of solutions with given metric importance.

\section{CONCLUSIONS}
In this paper, novel dynamic pricing algorithms are designed and evaluated for a novel EV CS pricing problem formulation. These algorithms are designed by integrating three components: (i) Bayesian Demand-Price model; (ii) Multi-Objective optimization algorithms NSGA II and NSGA III; and (iii) use of MCDM methods to select the solutions.

The performance of three-part dynamic pricing algorithms is validated by solving the EV CS pricing problem for two sites, JPL and Caltech.  This study also compares the three-part dynamic pricing algorithms with the widely used ToU pricing algorithm and SP.  Three widely used metrics are used to measure the quality of an EV CS pricing solution: total revenue, QoS, and PAR. The solution-finding capability of these algorithms when the importance of a metric is provided is also investigated.

We identify the better characteristics of the BM + NSGA III algorithm.  Also, the MCDM is seen to help find solutions given the metric's importance. Experiments also confirm the efficiency of BM + NSGA II and BM + NSGA III in terms of solution quality, i.e., in obtaining better solutions than ToU.  The results also reveal that BM + NSGA II and BM + NSGA III are more efficient in solving the EV CS pricing problem than the ToU pricing algorithm. 

The results can also guide decision-makers in making better CS pricing decisions when many CSs (in order of hundreds or more) need to be priced for efficiently charging at different CSs.  It is worth mentioning that the prices are obtained by escaping local best, exploring multiple solutions, and utilizing the strengths of different procedures like demand-price modelling, multi-objective optimization, and MCDM.


%Bibliography
% \bibliographystyle{unsrt}  
%\bibliographystyle{IEEEtran}
%\bibliography{IEEEabrv,bibilo} 

\end{document}